\documentclass{amsart}
\usepackage{slashed}
\usepackage{amssymb}
\usepackage[footnotesize,bf]{caption}

\theoremstyle{definition}

\theoremstyle{remark}

\def\beq{\begin{eqnarray}}
\def\eeq{\end{eqnarray}}
\def\bsp{\begin{split}}
\def\esp{\end{split}}

\def\d{\mathrm{d}}

\newcommand{\be}{\begin{equation}}
\newcommand{\ee}{\end{equation}}

\newcommand{\nnb}{\nonumber}
\newcommand{\p}{\partial}

\def \hbm #1 {\mbox{\boldmath{$\hat m^{(#1)}$}}}

\newcommand{\ol}{\overline}

\begin{document}

\hspace{8.5cm} NIKHEF/2007-008
\vspace{1cm}
\title{\textbf{Vanishing scalar invariant spacetimes in supergravity}}
\author{\textbf{A. Coley, A. Fuster, S. Hervik and N. Pelavas}}

\address{Department of Mathematics and Statistics,
Dalhousie University, Halifax, Nova Scotia, Canada B3H 3J5 (AC, SH
and NP); National Institute for Nuclear and High-Energy Physics
(NIKHEF), Kruislaan 409, 1098 SJ, Amsterdam, The Netherlands (AF)}
\email{aac,herviks,pelavas@mathstat.dal.ca;~fuster@nikhef.nl}

\date{\today}
\maketitle

\begin{abstract}

We show that the higher-dimensional vanishing scalar invariant
(VSI) spacetimes with fluxes and dilaton are solutions of type IIB
supergravity, and we argue that they
are exact solutions in string theory.
We also discuss the supersymmetry properties of VSI 
spacetimes.

\end{abstract}
\vspace{0.5cm}
\noindent
[PACS: 04.20.Jb, 04.65.+e]

\vskip .1in

\section{Introduction}

VSI spacetimes are $N$-dimensional Lorentzian spacetimes in which
all curvature invariants of all orders vanish  \cite{Higher}.
Recently, we presented all of the metrics for the higher-dimensional 
VSI spacetimes, which can be of Ricci type N or III \cite{CFHP}. 
The Ricci type N VSI spacetimes include the  higher-dimensional (generalized)
pp-wave spacetimes, which have been the most studied in the
literature and are known to be exact solutions of supergravity and
in string theory. However, many of the mathematical properties of
VSI  spacetimes (in general, and the pp-wave spacetimes in
particular) may not be familiar.

In this paper we will show
that all Ricci type N VSI spacetimes are solutions of supergravity
(and argue that Ricci type III VSI spacetimes are also
supergravity solutions if supported by appropriate sources). The VSI Ricci type III 
supergravity solutions to be presented are new. We also find some new
Ricci type N supergravity solutions. We
explicitly study type IIB supergravity, but similar
results are expected in all supergravity theories. We also argue
that, in general, the VSI spacetimes are exact string solutions to all orders
in the string tension.

We then discuss which VSI supergravity spacetimes can admit
supersymmetry. It is known that in general if a spacetime admits a
Killing spinor, it necessarily admits a null or timelike Killing
vector. Therefore, a necessary (but not sufficient) condition for
a particular supergravity solution to preserve some supersymmetry
is that the spacetime possess such a Killing vector. In the
Appendix we prove that VSI spacetimes whose metric functions have
dependence on the light-cone coordinate $v$ cannot possess a null
or timelike Killing vector. Hence, only VSI spacetimes with a
covariantly constant null vector are candidates to preserve
supersymmetry. Such spacetimes include not only pp-waves but also
(the more general) spacetimes of algebraic Weyl type III(a).

We therefore study the supersymmetry properties of VSI type IIB
supergravity solutions with a covariantly constant null vector. We focus on solutions 
of Weyl type
III(a), since type N spacetimes have been studied extensively. We conclude there 
are no such supersymmetric solutions in the vacuum type III(a) case. We
present explicit examples of Weyl type III(a) 
NS-NS (one-half) supersymmetric solutions.

\subsection{Higher-dimensional VSI metrics}

All curvature invariants of all orders vanish in an
$N$-dimensional Lorentzian (VSI) spacetime if and only if there
exists an aligned shear-free, non-expanding, non-twisting,
geodesic null direction $\ell^a$ along which the Riemann tensor
has negative boost order \cite{Higher}. The VSI spacetimes can
be classified according to their Weyl type, Ricci type and the
vanishing or non-vanishing of the `spin coefficient' $\varepsilon$
\cite{class}. In \cite{CFHP}, the explicit metric forms for higher
dimensional VSI spacetimes were presented.

Since VSI spacetimes possess a null vector field $\ell$ obeying
\beq
\ell^{A}\ell_{B;A}=\ell^{A}_{~;A}=\ell^{A;B}\ell_{(A;B)}=\ell^{A;B}\ell_{[A;B]}=0; \label{null}
\eeq i.e., $\ell$ is geodesic, non-expanding, shear-free and
non-twisting, the VSI spacetimes belong to the higher-dimensional
{\em Kundt} class \cite{Higher}. It follows that any {VSI}
metric can be written in the form
\beq \d s^2=2\d u\left[\d
v+H(v,u,x^n)\d u+W_{i}(v,u,x^n)\d x^i\right]+\delta_{ij}\d x^i\d
x^j \label{Kundt}
\eeq
with $i,j=1,\dots,N-2$. The negative boost
order conditions of the Riemann tensor yield \cite{CFHP}

\beq W_{ i}(v,u,x^k)=v{W}_{ i}^{(1)}(u,x^k)+{W}_{ i}^{(0)}(u,x^k),
\label{Weq}\eeq \beq
H(v,u,x^k)=\frac{v^2}{8}({W}_i^{(1)})({W}^{(1)i})+v{H}^{(1)}(u,x^k)+{H}^{(0)}(u,x^k).
\label{Heq}\eeq The $W^{(1)}_i$ are subject to further
differential constraints: using the allowable freedom we can
choose \beq W^{(1)}_1 = -2\frac{\varepsilon}{x^1};\quad W^{(1)}_n
= 0,\;\; n =2, ..., N-2 \label{kin} \eeq (where $\varepsilon=0$
corresponds to $W^{(1)}_1 = 0$ and $\varepsilon=1$ corresponds to
$W^{(1)}_1 \neq 0$). Accordingly,
\beq
H(v,u,x^k)=\frac{v^2\varepsilon}{2(x^1)^2}
+v{H}^{(1)}(u,x^k)+{H}^{(0)}(u,x^k).
\eeq

We note that, in general, for the Kundt metrics there exist
coordinate transformations $x'^j=f^j(u,x^i)$ which can be used to
simplify \emph{either} the transverse metric \emph{or} the
functions  $W_{i}^{(0)}(u,x^k)$. Here we have used this to
eliminate the $u$-dependence in the transverse metric to get the
explicitly flat metric $\delta_{ij}\d x^i\d x^j$ (and hence in
these coordinates the $W_{i}$ are not zero). Under the remaining
allowable coordinate transformations we obtain $H^{(0)} (u, x^k)
\to H^{(0)} (u, x^k) - (h(u, x^k))_{,u}$, so that we can redefine
$W^{(0)}_i$ and essentially freely specify $H^{(0)}$ (e.g., we
could set $H^{(0)}$ to zero), and in the case $\varepsilon =0$, we
obtain $H^{(1)}{(u, x^k)} \to H^{(1)}(u, x^k) + G(u)$ (after
redefining $H^{(0)}$ and $W^{(0)}_i$), where $G(u)$ is freely
specifiable.

All of these spacetimes are VSI. The spacetimes above are in
general of Ricci and Weyl type III. Further progress can be made
by classifing the metric in terms of their Weyl type (III, N or O)
and their Ricci type (N or O) \cite{class}, and the form of
$\varepsilon$. The~metric functions $H$ and $W_{i}$ satisfy the
remaining Einstein equations. In Table $1$ in \cite{CFHP}, all of
the VSI spacetimes supported by appropriate bosonic fields
are presented and the metric functions are
listed\footnote{\; The Christoffel symbols needed for the
calculations are: In $\varepsilon=0$ VSI spacetimes,
$\Gamma_{\lambda v}^{\;\;\;\lambda}=\Gamma_{\lambda
i}^{\;\;\;\lambda}=0$ for any choice of $\lambda$. In
$\varepsilon=1$ VSI spacetimes, $\Gamma_{\lambda
v}^{\;\;\;\lambda}=0,\;\;\Gamma_{u 1}^{\;\;\;u}=-\Gamma_{v
1}^{\;\;\;v}=-\frac{1}{2}W_{1,v}$.}. It is the  higher-dimensional
(generalized) pp-wave spacetimes that have been most studied in
the literature. It is known that such spacetimes are exact
solutions in string theory \cite{string,hortseyt,tseytlin}, in
type-IIB superstrings with an ~R-R five-form \cite{matsaev}, also
with NS-NS form fields \cite{RT}. In higher dimensions, VSI
supergravity solutions can be constructed \cite{hortseyt}, and we
shall see that all VSI spacetimes are solutions of superstring
theory when supported by appropriate bosonic fields.

It is convenient to introduce the null frame \beq
\ell&=& \d u, \label{nframe1} \\
{\bf n}&=& \d v+H\d u+W_{ i}{{\bf m}}^{ i+1}, \label{nframe2} \\
{\bf m}^{ i+1}&=& \d x^i.  \label{nframe3} \eeq The Weyl tensor
can then be expressed as \cite{class} \beq C_{abcd} = 8 \Psi_{i}
\ell_{\{a} n_b \ell_{c} m^{i}_{d\}} + 8 \Psi_{ijk} m^{i}_{\{a}
m^{j}_b \ell_c m^{k}_{d\}} + 8 \Psi_{ij} \ell_{\{a} m^{i}_b
\ell_{c} m^{j}_{d\}} . \label{WeylIIIN} \eeq 
The case $\Psi_{ijk} \not= 0$ is of Weyl type III, while $\Psi_{ijk} = 0$ 
(and $\Psi_i \equiv 2\Psi_{ijj}=0$)
corresponds to type N. 
Further subclasses
of type III can be considered; for example, type III(a) where $\Psi_i=0$ but
$\Psi_{ijk} \neq 0$.
The Ricci
tensor is given by \beq R_{ab} =  \Phi \ell_{a}\ell_{b} + \Phi_{i}
(\ell_{a} m^{i}_{b} + \ell_{b} m^{i}_{a}). \label{ricci} \eeq The
Ricci type is $N$ if $\Phi_{i}=0 = R_{1i}$ (otherwise the Ricci
type is III; Ricci type O is vacuum). When $\varepsilon=0,1$ the
Ricci type N conditions $\Phi_{i}=0$ reduce to
\begin{eqnarray}
2H^{(1)}_{\quad,1} & = & \frac{2\varepsilon}{x^{1}}W^{(0)m}_{\quad\quad,m}-W^{(0)m}_{\quad\quad,m1}  \label{rn1} \\
2H^{(1)}_{\quad,n} & = & \Delta W^{(0)}_{n}-W^{(0)m}_{\quad\quad,mn}  \label{rn2}
\end{eqnarray}
subject to
\begin{equation}
\Delta W^{(0)}_{n,1}  =
\frac{2\varepsilon}{x^1}W^{(0)m}_{\quad\quad,mn}, ~~ \Delta
W^{(0)}_{m,n}  = \Delta W^{(0)}_{n,m},     \label{icrn2}
\end{equation}
where $\Delta=\partial^{i}\partial_{i}$ is the spatial Laplacian
and $m,n\geq 2$. As a result, in Ricci type N spacetimes
$H^{(1)}$ can be determined as a
function of the $W^{(0)}_i$ (in contrast to the
Ricci type III case).

For the VSI spacetimes, the aligned, repeated, null vector $\ell$
is a null Killing vector (KV) if $\ell_{1;1}=0=\ell_{(1;i)}$
(i.e., $\varepsilon=0$), whence $H_{,v}=0$ and $W_{i,v}=0$ and the
metric no longer has any $v$ dependence. Furthermore, since
$L_{AB}:=\ell_{A;B}=\ell_{(A;B)}$ it follows that in this case if
$\ell$ is a null KV then it is also covariantly constant. In
general, the higher-dimensional VSI metrics admitting a null KV
(and hence a covariantly constant null vector (CCNV)) are of Ricci
and Weyl type III \cite{CFHP}. The subclass of Ricci type N CCNV
spacetimes are related to the ($F=1$) chiral null models of
\cite{hortseyt}. The subclass of Ricci type N and Weyl type III(a)
spacetimes includes the relativistic gyratons \cite{frolov}. The
subclass of Ricci type N and Weyl type N spacetimes are the
generalized pp-wave spacetimes.

\section{VSI spacetimes in IIB supergravity}

Our aim is to construct bosonic solutions of IIB supergravity for
which the spacetime is  VSI. We consider solutions with non-zero
dilaton, Kalb-Ramond field and RR $5$-form. The corresponding field
equations\footnote{\;The equations of motion of IIB supergravity
are  given, for example, in \cite{iibeqs}.} are \beq
R_{\mu\nu}-\frac{1}{2}R\;g_{\mu\nu}&=&-2\nabla_{\mu}\p_{\nu}\phi+\frac{1}{4}H_{\mu\lambda\rho}H_{\nu}^{\;\;\lambda\rho}+e^{2\phi}\frac{1}{4\cdot4!}F_{\mu\lambda\rho\kappa\sigma}F_{\nu}^{\;\;\lambda\rho\kappa\sigma} \label{eqR} \\
0&=&F_{ijklm}H^{klm} \label{eqFH} \\
\nabla_{\mu}\p^{\mu}\phi&=&-\frac{1}{4}R+\frac{1}{4\cdot 2\cdot 3!}H^2+\p_k\phi\p^k \phi  \label{eq0} \\
0&=&\nabla_{\lambda}H^{\lambda \mu \nu}-2(\p_{\lambda}\phi)H^{\lambda \mu \nu} \label{eqKalb-Ramond field1} \\
H&=&dB \label{hexact} \\
dF&=&0 \label{fexact} \\
F&=&*F
\eeq

The VSI requirement implies that all curvature tensors must be of
negative boost order. It is therefore reasonable that we similarly
require that the quadratic terms in $H$ and $F$ in eqn.
(\ref{eqR}) are also of negative boost order. Since $H$ and $F$
are forms (hence, antisymmetric), we must have
$F=(F)_{-1}+(F)_0+(F)_1$ (similarly for $H$), where $(~)_b$ means
projection onto the boost-weight $b$ components. As an example,
consider $(F)_1$ which can be written
$(F_{\mu\lambda\rho\kappa\sigma})_1=n_{[\mu}\xi_{\lambda\rho\kappa\sigma]}$
where $\xi_{\lambda\rho\kappa\sigma}$ is a four-form and
$\xi_{\lambda\rho\kappa\sigma}n^{\lambda}=\xi_{\lambda\rho\kappa\sigma}\ell^{\lambda}=0$.
We then have the boost-weight $2$ component (analogously for $H$):
\[ \left(F_{\mu\lambda\rho\kappa\sigma}F_{\nu}^{\;\;\lambda\rho\kappa\sigma}\right)_2=\frac 1{25} n_{\mu}n_{\nu}\xi_{\lambda\rho\kappa\sigma}\xi^{\lambda\rho\kappa\sigma}. \]
The factor
$\xi_{\lambda\rho\kappa\sigma}\xi^{\lambda\rho\kappa\sigma}$ is a
sum of squares, so that requiring that boost-weight $2$ components
of the $H^2+F^2$ terms should vanish thus implies that
$\xi_{\lambda\rho\kappa\sigma}=0$, and hence, $(F)_1=0$. A similar
calculation of the boost-weight $0$ components enables us to show
that if the quadratic terms of eqn. (\ref{eqR}) only have negative
boost weight terms then $F$ and $H$ only possess negative boost
weight terms\footnote{\; One can, in principle, imagine a very
special (and unnatural) situation where the derivatives of $\phi$
exactly cancel the non-negative boost-weight terms of $H^2+F^2$;
however, we shall not consider this possibility further here.}.
Assuming, in addition, that $\nabla_{\mu}\phi$ has negative boost
order, we thus have:
\[ \nabla\phi=\left(\nabla\phi\right)_{-1}, \quad H=\left(H\right)_{-1}, \quad F=\left(F\right)_{-1}. \]
Note that this immediately implies that eqns. (\ref{eqFH}) and
(\ref{eq0}) are satisfied. Furthermore, this also means that the
forms can be written \be
H_{\mu\nu\rho}=\ell_{[\mu}\tilde{B}_{\nu\rho]},~F_{\mu\lambda\rho\kappa\sigma}=\ell_{[\mu}\varphi_{\lambda\rho\kappa\sigma]}
\ee and
$\ell^{\mu}\tilde{B}_{\mu\nu}=\ell^{\mu}\varphi_{\mu\rho\kappa\sigma}=0$,
$n^{\mu}\tilde{B}_{\mu\nu}=n^{\mu}\varphi_{\mu\rho\kappa\sigma}=0$;
i.e., $\tilde{B}$ and $\varphi$ only have transverse components.
By calculating $\d \phi$ and requiring this only possess negative
boost order terms, we obtain $\phi=\phi(u)$.

\subsection{Ricci type N solutions}

We first construct solutions with Ricci type N VSI spacetimes. We
postulate the following ansatz motivated by the preceding argument \be
g_{\mu\nu}=g_{\mu\nu}^{\mbox{\tiny VSI}},\;\phi= \left\{
\begin{array}{cc} \phi(u) & (\varepsilon=0)
\\ \phi_0 & (\varepsilon=1) \end{array}
\right\},\;H_{\mu\nu\rho}=\frac{1}{4}\ell_{[\mu}\tilde{B}_{\nu\rho]},\;F_{\mu\lambda\rho\kappa\sigma}=\ell_{[\mu}\varphi_{\lambda\rho\kappa\sigma]
} \label{ansatz}
\end{equation}
where $\phi_0$ is a constant, $\tilde{B}$ and $\varphi$ are a two-
and four-forms with no dependence on $v$, and $\ell$ is the null
vector field in (\ref{null}). Eqns. (\ref{eqFH}), (\ref{eq0}) are
automatically satisfied. From eqns. (\ref{hexact}), (\ref{fexact}), 
$\tilde{B}_{\nu\rho}=\tilde{B}_{\nu,\rho}-\tilde{B}_{\rho,\nu}$ 
\footnote{\;Note that the relationship between $B$, as defined  in (\ref{hexact}),
 and $\tilde{B}$ is $B_{\nu\rho}=\ell_{[\nu}\tilde{B}_{\rho]}$.} and $\varphi$ has to be a closed form. We are left with the
following equations
%i've multiplied the einstein eq. by -1
\begin{eqnarray}
 x^1 \triangle \left( \frac{H^{(0)}}{x^1} \right) &+&\left( \frac{W^{(0)m}W^{(0)}_m}{x^1}\right),_{1}
-2H^{(1)},_{m} W^{(0)m} -H^{(1)} W^{(0)m},_{m}
\nonumber \\  & - &\frac{1}{4}W_{mn}W^{mn}- W^{(0)m}_{\quad\quad,mu}=-\frac{1}{4\cdot4}\tilde{B}_{ij}\tilde{B}^{ij}-3!\;e^{2\phi_0}\varphi^2\;\;(\varepsilon=1)  \label{Ruueps1}
\end{eqnarray}
%[note: the dilaton has to be constant in the $\varepsilon=1$ case because otherw%ise there is a term containing $v$ on the r.h.s. of the above equation; there %would also be a contribution to $R_{u1}$ (see Appendix)]
\begin{eqnarray}
\triangle H^{(0)}&- &\frac{1}{4}W_{mn}W^{mn}
-2H^{(1)},_{m} W^{(0)m} -H^{(1)} W^{(0)m},_{m}
\nonumber \\  & - & W^{(0)m}_{\quad\quad,mu}=2\phi''+H^{(1)}\phi'-\frac{1}{4\cdot4}\tilde{B}_{ij}\tilde{B}^{ij}-3!\;e^{2\phi}\varphi^2
\;\;(\varepsilon=0) \label{Ruueps0}
\end{eqnarray}
\be
\p_i \tilde{B}^{ij}=0 \label{eqKalb-Ramondfield2}
\ee
\be
\varphi=*_8\varphi
\ee
In the above equations $m,n=2,\ldots,8$ and $i,j=1,\ldots,8$. Prime denotes derivative with respect to $u$ and $*_8$ is the Hodge operator in the eight-dimensional transverse space. \\

The solutions above are of Weyl type III. They contain previously
known solutions such as the string gyratons \cite{strgyr} and pp-wave 
supergravity solutions \cite{matsaev,hortseyt,tseytlin,RT}. The latter arise in the Weyl type N limit of the $\varepsilon=0$ solutions (see \cite{CFHP}). \\

\subsection{Ricci type III solutions}

Ricci type III VSI spacetimes exist if appropriate source fields can be found. Recall that such a Ricci tensor must have boost weight $-1$ components. We note that for a general tensor product
\[ (T\otimes S)_b=\sum_{b=b'+b''}(T)_{b'}\otimes(S)_{b''} \]
Hence, the quadratic terms in $H$ and $F$ in eqn. (\ref{eqR})
necessarily have boost weight $-2$. Projection of eqn. (\ref{eqR})
onto boost weight $-1$ components then gives
\[ \left(R_{\mu\nu}\right)_{-1}=-2\left(\nabla_{\mu}\p_{\nu}\phi\right)_{-1}\]
We conclude that the space can only be of Ricci type III if
$\left(\nabla_{\mu}\p_{\nu}\phi\right)_{-1}$ is non-zero. This
implies that $\varepsilon=1$ and $\phi=\phi(u)$. Therefore, there
are no Ricci III supergravity solutions where $\ell$ is a
covariantly constant null vector (CCNV)
\footnote{\; This is in the context of type IIB supergravity (with the sources
considered) and the conditions described above.
However, we cannot exclude the existence of supersymmetric
Ricci type III solutions in more general situations.}. 
This is perhaps unfortunate, as such solutions would have been good candidates to preserve supersymmetry. \\

Motivated by the preceding argument we construct a solution with
non-constant dilaton $\phi=\phi(u)$ in the $\varepsilon=1$ case.
Eqns. (\ref{eqR}) read
\begin{eqnarray}
x^1 \triangle \left( \frac{H^{(0)}}{x^1} \right) &+& \left( \frac{W^{(0)m}W^{(0)}_m}{x^1}\right),_{1}
-2H^{(1)},_{m} W^{(0)m} -H^{(1)} W^{(0)m},_{m}- \frac{1}{4}W_{mn}W^{mn}
\nonumber \\  &-& W^{(0)m}_{\quad\quad,mu}+\frac{2v}{(x^1)^2}\phi'=2\phi''+2\left(H^{(1)}+\frac{v}{(x^1)^2}\right)\phi' \label{Ruueps1dil}
\eeq
%The equations for $R_{u1}$, $R_{uj}$, $j=2,...,8$
\beq
H^{(1)}_{\quad,1} & = & \frac{1}{x^{1}}W^{(0)m}_{\quad\quad,m}-\frac{1}{2}W^{(0)m}_{\quad\quad,m1} +\frac{2}{x^1}\phi' \label{eq1} \\
2H^{(1)}_{\quad,n} & = & \Delta W^{(0)}_{n}-W^{(0)m}_{\quad\quad,mn} \label{eq2}
\eeq
Note that the $v$-dependent terms in (\ref{Ruueps1dil}) cancel each other. On the other hand, we can determine $H^{(1)}$ from (\ref{eq1}), (\ref{eq2}). The complete metric function is
\be
H  =  H^{(0)}(u,x^{i})+\frac{1}{2}\left(\tilde{F}-W^{(0)m}_{\quad\quad,m}\right)v+\frac{v^2}{2(x^1)^2}, \label{e1hIII}
\ee
where $\tilde{F}=\tilde{F}(u,x^i)$ is a function satisfying
\begin{equation}
\tilde{F},_1=\frac{2}{x^1}(W^{(0)m}_{\quad\quad,m}+2\phi'),\;\;\;\;\tilde{F},_n
=\Delta W^{(0)}_{n}.  \label{Feps1}
\end{equation}
The solution above is, to our knowledge, the first supergravity solution of Ricci type III.
The dilaton dependence on $u$ is crucial to construct the solution, and it reduces
to Ricci type N when the dilaton is constant (or absent). VSI supergravity solutions of
Ricci type III with form fields only do not exist. However, the solution above can be generalized in a straightforward way to include the form fields in (\ref{ansatz}). \\

The Ricci type N, Weyl type III solutions in the previous section
can be reduced to Weyl type N. On the contrary, the Ricci type III
solution presented here can only have Weyl type III
\footnote{\;Requiring the solution to have vanishing boost weight
$-1$ Weyl components reduces it to the Ricci and Weyl type N
solution.}.

\subsection{Solutions with non-zero $\mathbf{F_1}$, $\mathbf{F_3}$.}
The above solutions can be generalized to include non-zero $F_1$, $F_3$ RR fields. It is well-known that $SL(2,{\mathbb{R}})$ is the classical S-duality symmetry group for IIB supergravity (see, for example, \cite{TOrtin}). Such a transformation can be parametrized by \cite{hassan}
\be
S=\left(
\begin{array}{cc} p & q
\\ r & s \end{array}\right) \label{trafo}
\ee
with $ps-qr=1$. The fields transform according to
\be
 \left(
\begin{array}{c} A'_2
\\ B' \end{array}\right)=\left(
\begin{array}{cc} p & q
\\ r & s \end{array}\right)\left(
\begin{array}{c} A_2
\\ B \end{array}\right),\;\;\;\;\tau'= \frac{p\tau+q}{r\tau+s}
\ee
where $A_2$ and $B$ are the RR $2$-potential and the Kalb-Ramond field,
and $\tau=A_0+ie^{-\phi}$, where $A_0$ is the RR scalar and $\phi$ is the dilaton. The metric and RR $5$-form remain invariant\footnote{\;The S-duality symmetry becomes manifest when the metric is in the Einstein frame: $\hat{g}=e^{\phi/2}g$. Such a rescaling does not affect the character of the VSI solutions presented.}. The VSI solutions presented have $A_0=A_2=0$. Under a transformation (\ref{trafo})
\be
A'_2=qB,\;\;\;B'=sB
\ee  
In this way one can generate a non-zero $F'_3$ which is proportional to $H$; the Kalb-Ramond field gets rescaled. The dilaton and RR scalar can be read from
\beq
sA'_0-e^{-(\phi+\phi')}r&=&q \\
rA'_0+se^{(\phi-\phi')}s&=&p
\eeq 
For solutions with $\phi=\phi(u)$ one obtains $(F'_1)_u=\partial_uA'_0$. For solutions with a constant dilaton the RR $1$-form remains zero. \\

\subsection{String corrections}

In four dimensions VSI spacetimes are known to be exact string
solutions to all orders in the string tension $\alpha'$ even in the presence
of additional fields \cite{coley}. Using the arguments of \cite{string}, 
higher-dimensional supergravity solutions supported by appropriate fields
(e.g., with the dilaton and Kalb-Ramond field and appropriate form fields) are
also known to be exact solutions in string theory
\cite{matsaev,hortseyt,tseytlin,RT}. Similarly \cite{coley}, it
can be argued that the VSI supergravity spacetimes are exact string
solutions to all orders in the string tension $\alpha'$, 
at least in the presence of a dilaton and Kalb-Ramond field. Higher-dimensional 
pp-waves are also exact solutions of string theory with RR fields (e.g., a $F_5$ field);
it is to be expected, from an analysis of the perturbative counterterms, that this is also
the case for the special VSI supergravity solutions under consideration here.
Therefore, the VSI solutions presented may be of relevance in string theory.  
Note that these VSI
spacetimes are, in general, time-dependent string theory backgrounds.

\section{Supersymmetry}
Given a spinor $\epsilon$ on a Lorentzian manifold, the vector
constructed from its Dirac current \be k^a  = \bar\epsilon
\gamma^a \epsilon \ee is null or timelike. Moreover, if $\epsilon$
is a Killing spinor then $k^a$ is a Killing vector. This result
has been proven for a number of supergravity theories (for
example, $D=11$ \cite{hommth}, type IIB \cite{iibkv}), and it is
generally believed to hold in all theories of supergravity
(although the details may vary in each particular theory depending
on the specific field equations).
Therefore, a necessary (but not sufficient) condition for a particular 
supergravity solution to preserve some supersymmetry is that the 
involved spacetime possesses a null or timelike Killing vector. 

%The causal character of the Killing vector depends on the particular theory of superg%ravity considered. For example, in IIB supergravity it is always null \cite{iibkv}.
The existence of Killing vectors in VSI spacetimes in an arbitrary
number of dimensions has been studied. It is known that there can
exist no null or timelike Killing vector unless $\varepsilon =0$
and $H$ is independent of $v$ (see Appendix). Therefore,
there are no supersymmetric solutions in any other type of
supergravity theory. We therefore study only the supersymmetry
properties of VSI IIB supergravity solutions with a covariantly
constant null vector \cite{TOrtin}. These are of Ricci type N and
Weyl type III(a) or N \cite{CFHP}. We will focus on solutions of
Weyl type III(a), as the Weyl type N ones have been discussed
extensively in the literature.
We will consider two different cases: vacuum solutions and NS-NS solutions. 

The Killing spinor equation for pure gravitational solutions reads
\be
(\partial_{\mu}-\frac{1}{4}w_{\mu\;ab}\Gamma^{ab})\epsilon=0.
\ee Greek indices are curved indices $u,v,1,...,8$ and Latin
indices are tangent space indices $1,...,10$. The supersymmetry
parameter $\epsilon$ is a complex-valued $16$-component chiral
spinor. In our conventions, the basis of one-forms is: \beq
w^i&=&dx^i \nnb \\
w^9&=&du \nnb \\
w^{10}&=&dv+H\;du+W_i\;dx^i, \nnb \eeq and the corresponding
inverse frame is: \beq
e_i&=&\p_i -W_i\;\p_v \nnb \\
e_9&=&\p_u-H\;\p_v \nnb \\
e_{10}&=&\p_v. \nnb
\eeq
The components of the spin connection for Weyl III(a) VSI spacetimes are
\beq
w_{u\;ij}=\frac{1}{2}W_{ij},&\;\;&w_{u\;i9}=-H^{(0)},_i+W^{(0)}_{i,u} \\
w_{i\;j9}&=&\frac{1}{2}W_{ji}, \eeq It has been proved\footnote{\;
The analysis in \cite{mwaves} concerns five dimensions. However,
the result holds in higher dimensions so long as the transverse
space is flat \cite{pcom}.} that the only case where supersymmetry
arises is when \cite{mwaves} \be
\partial_k W_{ij}=0. \label{condsusy}
\ee
This condition is equivalent to the functions $W_i$ being linear in the transverse coordinates. In that case the spacetime reduces to Weyl type N \cite{CFHP}. \\

We discuss next supersymmetry on NS-NS solutions. The fermion supersymmetry transformations are given by:
\begin{eqnarray}
\delta \psi_{\mu}&=&(\partial_{\mu}-\frac{1}{4}\Omega_{\mu\;ab}\Gamma^{ab})\epsilon, \\
\delta \lambda&=& (\slashed{\partial}\phi-\frac{1}{6}\slashed{H})\epsilon,
\end{eqnarray}
where $\Omega$ is the torsionful spin connection
\be
\Omega_{\mu\;ab}=w_{\mu\;ab}+H_{\mu\;ab}.
\ee
In components we have
\beq
\Omega_{u\;ij}=\frac{1}{2}(W_{ij}+\tilde{B}_{ij}),&\;\;&\Omega_{u\;i9}=-H^{(0)},_i+W^{(0)}_{i,u} \\
\Omega_{i\;j9}&=&\frac{1}{2}W_{ji}, \eeq
%At first sight, the dilatino equation vanishes for an Kalb-Ramond fieldfield depending on%ly on $u$ ($\tilde{B}_{ij}=\tilde{B}_{ij}(u)$) if the dilaton and Kalb-Ramond fieldterm are appropriately% related. This would not constrain the Killing spinor in any way. However, thi%s would not modify condition (\ref{condsusy}) either and the resulting supersy%mmetric solutions would be of Weyl type N [ref?]. \\
%We can write the Killing spinor $\epsilon$ as
%\be
%\epsilon=\frac{1}{2}\Gamma^u\Gamma^v\epsilon+\frac{1}{2}\Gamma^v\Gamma^u\epsil%on=\epsilon_{+}+\epsilon_{-},
%\ee
%where $\epsilon_{+}$ ($\epsilon_{-}$) is $SO(8)$ positive (negative) chiral.
We consider solutions with at most one half of the supersymmetries
broken \footnote{\; In some cases, different fractions of
supersymmetry can be preserved \cite{rotbr}.}; i.e.,
$\Gamma^u\epsilon=0$. The dilatino variation then vanishes
automatically and the gravitino Killing equation reduces to
%we have to adjust the ansatz for the Kalb-Ramond fieldsuch that H_{u\;ij}=1/2\tilde{B}_{ij}
\be
\left(\partial_u-\frac{1}{4}(W_{ij}+\tilde{B}_{ij})\Gamma^{ij}\right)\epsilon=0,\;\;\;\partial_v\epsilon=\partial_i\epsilon=0.
\ee
We see that for consistency
\be
W_{ij}=\tilde{B}_{ji}+f_{ij}(u), \label{susycond}
\ee
where $f_{ij}$ are arbitrary functions of $u$. However, for our purposes we can take $f_{ij}=0$ since such functions are related to Weyl type N solutions. We therefore obtain eight (complex) constant Killing spinors and half of the supersymmetry is preserved. \\

From eqns. (\ref{susycond}) and (\ref{eqKalb-Ramondfield2}), the metric
functions $W_i$ satisfy \be \p_i \;(W_{i,j}-W_{j,i})=0.
\label{cond1} \ee This is a necessary condition for the spacetime
to be of Weyl type III(a). In addition it is required that \be
\p_k \;(W_{i,j}-W_{j,i}) \neq 0, \label{cond2} \ee The
supersymmetry analysis is similar to that of \cite{ssw} and the
resulting solutions are the IIB analogues to the supersymmetric
string waves. However, we have shown that the solutions can have a more general algebraic type than 
pp-waves. We present a few examples of such solutions below. \\

Consider the VSI metric\footnote{\; This metric is the ten-dimensional generalization of a five-dimensional VSI metric of Weyl type III(a) presented in \cite{CFHP}.}
\beq
W_1&=&0 \\
W_m&=&f_{mn}(u)x^nx^1 \\
H&=&H^{(0)}(u,x^i), \eeq where $f_{mn}$ are antisymmetric arbitrary
functions of $u$ and $m,n=2,...,8$. This spacetime satisfies
(\ref{cond1}), (\ref{cond2}) and is therefore of Weyl type III(a).
Supported by the dilaton and Kalb-Ramond field\be \phi=\phi(u),\;\;\;
\tilde{B}_{1m}=f_{mn}(u)x^n,\;\tilde{B}_{mn}=2f_{nm}(u)x^1, \ee
it is a supersymmetric solution of the type discussed above. The function $H^{(0)}$ can be determined from equation (\ref{Ruueps0}). \\

Another example involves the gyraton metric presented in \cite{frolov}. In ten dimensions
\beq
W_i&=&-\frac{\tilde{p}_i(u)x^{i+1}}{Q^4} \\
W_{i+1}&=&\frac{\tilde{p}_i(u)x^{i}}{Q^4} \\
H&=&H^{(0)}(u,x^i),
\end{eqnarray}
where $i$ only takes odd values $1,3,5,7$ and $Q=\delta_{jk}x^jx^{k}$, $j,k=1,...,8$. The $\tilde{p}_i$ are arbitrary functions
\footnote{\;The relation to the functions $p_i$ in \cite{frolov} is $p_1=\tilde{p}_1$, $p_2=\tilde{p}_3$, $p_3=\tilde{p}_5$, $p_4=\tilde{p}_7$.} of $u$. The gyraton metric satisfies (\ref{cond1}), (\ref{cond2}) and so its Weyl type is III(a). In \cite{strgyr} this spacetime was considered in the context of supergravity, together with a constant dilaton and Kalb-Ramond field of the form presented in the ansatz. As we have seen such solution can be generalized to include a dilaton depending on $u$. The (one-half) supersymmetric gyraton will be the one with
\be
\phi=\phi(u),\;\;\; \tilde{B}_{jk}=W_{kj}.
\ee
Again, $H^{(0)}$ can be determined from eqn. (\ref{Ruueps0}). This solution belongs to the class of {\it saturated} string gyratons in \cite{strgyr}. Another supersymmetric (AdS) gyraton solution is given in \cite{adsgyr}. \\

\section{Discussion}
We have constructed solutions of IIB supergravity with NS-NS and
RR fluxes and dilaton for which the spacetime has vanishing scalar
invariants (VSI). The solutions are classified according to their
Ricci type (N or III). The Ricci type N solutions are
generalizations of pp-wave type IIB supergravity solutions. The
Ricci type III solutions are characterized by a non-constant
dilaton field. The resulting spacetimes are summarized in
Table $1$.  Note that we have not attempted to 
classify the VSI supergravity solutions in terms of their holonomy.

The supergravity solutions of Ricci type III (with
$\phi=\phi(u)$) are new. In addition, although the results
presented above are explicitly for type IIB
supergravity,  similar results are expected in all supergravity
theories. We have also argued that the VSI spacetimes presented
are exact string solutions to all orders in $\alpha'$,
at least in the presence of a dilaton and Kalb-Ramond field.

\begin{table}[h] \label{table:1}
\begin{center}
\scriptsize{
\begin{tabular}{|c|c|c|}
\hline
&&\\
{\bf Ricci type}  & $\varepsilon=0$ & $\varepsilon=1$  \\
&&\\
\hline
&&\\

 III & none & $\phi=\phi(u)$, $H$, $F$ \\
&&\\
\hline
&&\\
 N & $\phi=\phi(u)$, $H$, $F$ & $\phi$ constant, $H$, $F$ \\

&& \\
\hline

\end{tabular}
} \caption{VSI supergravity solutions. $H$ and $F$ are given in
(\ref{ansatz}).}
\end{center}
\end{table}

We have also studied the supersymmetry properties of VSI
spacetimes. We have shown that VSI spacetimes whose metric
functions have dependence on the light-cone coordinate $v$ cannot
posses a null or timelike Killing vector. Hence, we have argued
that only VSI spacetimes with a covariantly constant null vector
are candidates to preserve supersymmetry. Such spacetimes include
not only pp-waves but also spacetimes of a more general algebraic
type, namely, spacetimes of Weyl type III(a). We have studied the
supersymmetries of vacuum and NS-NS Weyl type III(a) solutions.
The latter preserve one-half of the supersymmetry when the axion
and metric functions are appropriately related. We present two
explicit supersymmetric examples, one of them being the string
gyraton in \cite{strgyr}. It is likely that RR Weyl type III(a)
spacetimes preserve some supersymmetry as well; we will address
this question in future work.

\section{Appendix:  Killing vectors}

In this appendix we show that there can exist no null or timelike
Killing vectors in VSI spacetimes unless $\varepsilon =0$ and $H$
is independent of $v$ (i.e., $\frac{\partial} {\partial v}$ is a
covariantly constant null vector).

Writing the frame components of the Killing vector
  $\xi$ as  $\xi_{\ol{1}}  {\mbox{\boldmath $n$}} + \xi_{\ol{2}}
  {\mbox{\boldmath $\ell$}} +
  \xi_{\ol{i}}
  {\mbox{\boldmath $m$}}^{\ol{i}}$, the Killing equations
  become:
  \begin{eqnarray}
  && \xi_{\ol{1}, \ol{1}} = 0 \label{xi1} \\
  && 2 \xi_{(\ol{1}, \ol{2})}
   - \xi_{\ol{1}} H_{,v} - \xi_{\ol{i}}
   W_{i, v}\; \delta^{i \ol{i}} = 0  \label{xi2}  \\
   && \xi_{(\ol{1}, \ol{j})} = 0 \label{xi3}\\
   && \xi_{\ol{2}, \ol{2}} + \xi_{\ol{2}} H_{, v} +
   \xi_{\ol{i}}
   [H_{, i} - W_{i, u} + HW_{i, v}
   - H_{, v} W_i] \delta^{i \ol{i}} = 0 \label{xi4} \\
   &&2 \xi_{(\ol{2}, \ol{j})} - \xi_{\ol{1}} [ H_{,j} -
   W_{j, u} + HW_{j, v} - H_{, v} W_j]
   + \xi_{\ol{2}} W_{j, v} - \xi_{\ol{i}} \tilde{A}_{ij} \delta^{i\ol{i}} =
0
   \label{xi5}\\
   && \xi_{(\ol{i}, \ol{j})} = 0  \label{xi6}
   \end{eqnarray}
  where $\tilde{A}_{k\ell} \equiv
  2W _{[k,\ell]} - 2W_{[k, |v|} W_{\ell]}$
  (and is independent of $v$), and the
  directional derivatives are given by
  \begin{eqnarray}
  \partial_{\ol{1}} & = & \partial_v \nonumber \\
  \partial_{\ol{2}} & = & \partial_u - H\partial_v \nonumber \\
  \partial_{\ol{i}} & = & - W_i \partial_v + \partial_{x^i}
  \end{eqnarray}
  (and henceforth we shall write $\partial_{x^i} = \partial_i$ for
  simplicity).

  For the VSI metric  (\ref{Kundt}) the functions $H$ and $W_i$
  are given by (see (\ref{Weq}) and (\ref{Heq}))
  \begin{equation}
  H  =  \frac{1}{2} H^{(2)} v^2 + H^{(1)} v + H^{(0)}; \quad
  W_i = W_i^{(1)} v + W_i^{(0)},
  \end{equation}
where the functions $W_i^{(0)} , H^{(0)}, H^{(1)}$ depend on $(u,
x^k)$ (i.e., are independent of $v$).  There are two cases,
$\varepsilon =0$ and $\varepsilon =1$, which can be represented by
(see (\ref{kin}))
\begin{equation}
H^{(2)}  = \frac{\varepsilon}{(x)^2} , \; \;  x \equiv x^1
;\quad W_1^{(1)} = -\frac{2\varepsilon}{x}, \; \;
W_{m}^{(1)} = 0 \quad (m \neq 1).
\end{equation}

We can immediately integrate eqns. (\ref{xi1}) - (\ref{xi3}) to
obtain
\begin{eqnarray}
\xi_{\ol{1}} & = & \xi(u, x^k)\\
\xi_{\ol{2}} & = & \frac{1}{2} \{\xi H^{(2)} - \xi_{,j} W_j^{(1)}
\}
v^2\\
&& + \{\xi H^{(1)} + L_j W_j^{(1)} -
\xi_{, u} \}v + \eta(u, x^k)\nonumber\\
\xi_{\ol{i}} & = & - \xi_{, i} v + L_i (u, x^k)
\end{eqnarray}
where $\xi, \eta$ and $L_i$ are arbitrary functions of $(u, x^k)$
  (and the repeated index $j$ indicates summation).  The
remaining eqns. (\ref{xi4})-(\ref{xi6}) become polynomials in $v$
of order $\mathcal {O}(v^3), \mathcal {O}(v^2), \mathcal {O}(v)$,
respectively, which can be solved to each power of $v$ separately.

Setting the $\mathcal {O}(v)$ term in eqn. (\ref{xi6}) to zero, we
obtain
\begin{eqnarray}
\varepsilon = 0: && \xi = f_{m} (u) x^{m}
+ f_1(u) x + g(u)\nonumber\\
\varepsilon = 1: && \xi = \frac{1}{x} f_{m} (u)
x^{m} + \frac{1}{x} f_1 (u) + g(u)
\end{eqnarray}
where $m = 2, \ldots, N - 2$ (i.e., $m \neq
1$), whence it follows that the $\mathcal {O}(v^3)$ and $\mathcal
{O}(v^2)$ terms in eqns. (\ref{xi4}) and (\ref{xi5}),
respectively, now vanish. Setting the $\mathcal {O}(v^2)$ term in
eqn. (\ref{xi4}) to zero and the $\mathcal  {O}(v)$ term in eqn.
(\ref{xi5}) to zero, respectively, we then obtain
\begin{equation}
  \frac{\varepsilon^2}{x^2} \{g'- \xi_{,u} + H^{(1)} (\xi -g)
- \frac{\varepsilon}{x}L_1   + \xi_{,j} W_j^{(0)}  \}
  - \xi_{,i} H^{(1)}_{,i}=0
\end{equation}
and
\begin{equation}
\xi_{,k} (W_{k,j}^{(0)} - W^{(0)}_{j,k}) + 2 \xi_{,j} H^{(1)} - 2
\xi_{,ju} + \; \xi_{,k} W_k^{(0)} W_j^{(1)} + (L_k W_k\;
^{(1)})_{,j} = 0. \label{stuff}
\end{equation}

\subsubsection{The case $\varepsilon =1$}

Let us assume that there exists a nontrivial solution to the
timelike or null Killing equations.  Rotating the frame we can
then align $\xi_{i}$; i.e., we can use the remaining frame freedom
to choose $\xi_{\ol{i}} = \xi^I \delta^I_i$, where $\xi^I = -
\xi_{, I} v +L$ and $\xi_{,J} = 0$ ($J \neq I$; i.e. $\xi = g(u,
x^I)$).  In this case we can then integrate eqn. (\ref{xi3}) to
obtain $\xi_{,I}=0$, so that $\xi =g(u)$, and consequently $L_{,i}
=0$, so that $L = \ell(u)$. Finally, eqn. (\ref{xi6}) implies that
$\ell(u) = 0$ or $I \neq 1$, whence eqn. (\ref{xi5}) is satisfied
identically. Hence, we have that
\begin{eqnarray}
   \xi_{\ol{1}}  &\equiv &  g(u), \quad \xi_{\ol{2}} =
\frac{1}{2x^2} gv^2 + (gH^{(1)} - g') v+
\eta (u, x^k), \nonumber \\
    \xi_{\ol{i}} & = & \ell(u) \delta^I_{\ol{i}} \; (I \neq 1)
    \label{star}
\end{eqnarray}
The remaining eqns. to be solved become
\begin{eqnarray}
&& \{\frac{1}{x^2} (\eta - gH^{(0)})  + (gH^{(1)} - g')_{,u} +
\ell
[H^{(1)}_{,I} - \frac{1}{x^2} W^{(0)}_I] \} v \nonumber\\
&& \qquad  + \{H^{(1)} \eta + \eta_{, u} - (gH^{(1)} - g')H^{(0)}
+ \ell [H^{(0)}_{, I} - W^{(0)}_{I,u} - H^{(1)} W^{(0)}_I]\} =0,
\label{A}
\end{eqnarray}
and
\begin{eqnarray}
&&\{\eta_{,j} - \frac{2}{x} \eta \delta^1_j + \ell^\prime
\delta^I_j + g' W^{(0)} _j + g(W^{(0)}_{j,u} - H^{(0)}_{,j}
+ \frac{2}{x} \delta^1_j H^{(0)})\nonumber\\
&& \qquad -\ell [W^{(0)}_{I,j} - W^{(0)}_{j,I} - \frac{2}{x}
\delta^1_j W^{(0)}_{I}] \}= 0. \label{B}
\end{eqnarray}
Setting the $\mathcal  {O}(v)$ term in eqn. (\ref{A}) to zero we
obtain
\begin{equation}
\eta = gH^{(0)} - x^2 (gH^{(1)} - g^\prime)_{,u} + \ell [W^{(0)}_I
- x^2 H^{(1)}_{,I}],
          \label{C}
\end{equation}
whence the remaining eqns. yield the contraints
\begin{eqnarray}
&&2g' H^{(0)} + gH^{(0)}_{,u} - x^2 \{(gH^{(1)} - g')_{, uu}
+ H^{(1)} (gH^{(1)} - g')_{,u}\} \nonumber\\
&& \qquad + \ell^\prime [W^{(0)}_I - x^2 H^{(1)}_{,I}] + \ell [
H^{(0)}_{,I} - x^2 \{H^{(1)} H^{(1)}_{,I} + H^{(1)}_{,Iu}\}] = 0,
\label{D}
\end{eqnarray}
and
\begin{equation}
\{g[W^{(0)} _j - x^2 H^{(1)}_{,j}]\} {_{,u}} + \ell^\prime
\delta^I_j + \ell\{W^{(0)}_j - x^2 H^{(1)}_{,j}\}_{,I} = 0.
   \label{E}
\end{equation}
For nontrivial functions $g$ and $\ell$, these equations reduce to
constraints on the metric functions $W^{(0)}_i, H^{(1)}$ and
$H^{(0)}$.

\subsubsection{The case $\varepsilon =0$}

In this case, $H^{(2)} = 0, W^{(1)}_i = 0$, and
\begin{eqnarray*}
\xi_{\ol{1}} & = & f_{\hat{i}}(u) x^{\hat{i}} +
f_1 (u) x + g(u) = \xi(u, x^u)\\
\xi_{\ol{2}} & = & (\xi H^{(1)} - \xi_{,u}) v + \eta (u, x^k)\\
\xi_{\ol{i}} & = & -f_i v + L_i
\end{eqnarray*}
The remaining Killing eqns. become (summing over $i$)
\begin{eqnarray}
&& \{-f_i H^{(1)}_{,i}\}v^2 + \{(\xi H^{(1)} - \xi_{, u})_{,u} +
L_i H^{(1)}_{,i} - f_i [H^{(0)}_{,i} - W^{(0)}_{i,u} -
H^{(1)} W^{(0)}_{i}]\}v \nonumber\\
&& \qquad + \{\eta_{, u} + H^{(1)} \eta - H^{(0)} (\xi H^{(1)} -
\xi_{,u}) + L_i [H^{(0)}_{,i} - W^{(0)}_{i,u} - H^{(1)} W^{(0)}_i
] \} = 0, \label{e1}
\end{eqnarray}
and
\begin{eqnarray}
  && \{-2f_{j,u} + 2f_i W^{(0)}_{[i,j]}\}v
  + \{\eta_{,j} + H^{(0)} f_j + L_{j,u} + W^{(0)}_j \xi_{,u}\nonumber\\
&& \qquad -\xi[H^{(0)}_{,j} - W^{(0)}_{j,u} ]
  - 2L_i W^{(0)}_{[i,j]} \} = 0.
  \label{e2}
\end{eqnarray}

The Killing vector $\frac{\partial}{\partial v}$ corresponds to
the solution $f_i = L_i = 0, \xi =0, \eta = \eta_0$ (with $H^{(1)}
= 0$) in eqns. (\ref{e1})  and (\ref{e2}).  Notice that there is
second solution to these eqns. (corresponding to $g'=0$) with $f_i
= L_i =0$ when $\xi = \xi_0$ and $\eta = H^{(0)} \xi_0$ (where
$\xi_0$ is a constant), provided that $ H^{(0)}_{,u} =
W^{(0)}_{j,u}  = 0  = H^{(1)}_{,u}; $ hence this corresponds to
the case in which all of the metric functions are independent of
$u$.  In this case $\xi_{\ol{1}} = \xi_0$ and $\xi_{\ol{2}} = H
\xi_0$, so that $\xi  = \xi_0 (dv + 2Hdu + W_i dx^{i+1})$, and the
corresponding Killing vector is $ \xi_0 \frac{\partial}{\partial
u}$, as expected.   Note that $|\frac{\partial}{\partial u}|^2 =
2H$, and so this Killing vector is timelike or null only when $H
\leq 0$.

\subsection{Timelike and null Killing vectors}

By direct calculation we find that
\begin{eqnarray}
|{\mbox{\boldmath $\xi$}}|^2 & = & g^{\ol{a} \ol{b}}
\xi_{\ol{a}} \xi_{\ol{b}}\nonumber\\
& = & 2 \xi_{\ol{1}} \xi_{\ol{2}} + \delta^{\ol{i} \ol{j}}
\xi_{\ol{i}} \xi_{\ol{j}}. \label{TagTLKV}
\end{eqnarray}
Assuming the Killing vector is null or timelike, we obtain
\begin{equation}
\{(\frac{\varepsilon \xi}{x} + \xi_{,1})^2 + \xi_{, \hat{i}}
\xi_{,\hat{i}}\} v^2 + 2\{\xi(\xi H^{(1)} - \frac{2
\varepsilon}{x} L_1 - \xi_{,u}) - \xi_{, i} L_i\}v + \{2 \eta \xi
+ L_i L_i\} \leq 0 \label{TagKVxi}
\end{equation}
for all coordinate values in the local chart.   In particular,
this is satisfied for all values of $v$ (positive and negative).
Hence, it follows that
\begin{equation}
\left( \frac{\varepsilon \xi}{x} + \xi_{,1} \right)^2 + \xi_{,
\hat{i}} \xi_{,\hat{i}} = 0,
\end{equation}
which, since the left-hand-side is the sum of positive-definite
terms, implies that
\begin{equation}
\frac{\varepsilon\xi}{x} + \xi_{,1} = 0; \quad \xi_{, \hat{i}} =
0. \label{TagA}
\end{equation}
Hence, since (\ref{TagKVxi}) is satisfied for both positive and
negative values of $v$, we must have that
\begin{equation}
\xi \left( \xi H^{(1)} - \frac{2 \varepsilon}{x} L_1 - \xi_{, u}
\right)
  - \xi_{, 1} L_1 = 0, \label{TagB}
\end{equation}
and consequently
\begin{equation}
2 \eta \xi + L_i L_i \leq 0. \label{TagC}
\end{equation}

\subsubsection{The case $\varepsilon =1$}  From eqn. (\ref{star}), we have that
$\xi_i = \xi \delta^I_i, \xi = g(u)$ and $L = \ell(u)$. It
immediately follows from (\ref{TagA}) that
$$ \varepsilon \xi = 0. $$
(Assuming $\varepsilon \neq 0$) we then have that $\xi = 0$, and
hence from eqn. (\ref{TagC}), $L=0$. Eqn. (\ref{TagTLKV}) then
implies $\eta = 0$, and consequently in this case we can only
obtain the trivial solution (i.e., {\em it there are no timelike
or null Killing vectors for $\varepsilon =1$}).

\subsubsection{The case $\varepsilon =0$} From eqn. (\ref{TagA}) it follows that
$$ \xi = g(u); \quad (f_i = 0)$$
whence eqn. (\ref{TagB}) implies that
$$ g(gH^{(1)} - g') = 0 $$
(and $\xi_{\ol{2}} = \eta (u, x^k), \xi_{\ol{i}} = L_{{i}}$; i.e.,
the components of the Killing vector have no $v$ dependence).

If $g =0$, then (\ref{TagC}) implies that $L_i =0$, whence eqns.
(\ref{e1}) and (\ref{e2}) yield $\eta_{,j} = 0$, so that $\eta =
\eta(u)$, and $\eta' + H^{(1)} \eta = 0$.  In the non-trivial case
$(\eta \neq 0)$, this implies that $H^{(1)} = H^{(1)} (u)$.  If $g
=0$, we have that $gH^{(1)} - g'$, whence again $H^{(1)}
  = H^{(1)} (u)$.
  In either case, we can always effect a coordinate transformation
  to set $H^{(1)} = 0$.  In this case $H$ has
  no $v$ dependence, and {\it hence the spacetime admits a covariant
  constant null vector}
$\frac{\partial}{\partial v}$ \cite{CFHP,TOrtin}; that is, eqns.
(\ref{e1}), (\ref{e2}) and (\ref{TagC}) always admit the solution
$g = 0, L_j = 0$, $\eta = \eta_0$ (a constant), corresponding to
the null Killing vector $\frac{\partial}{\partial v}$ (and there
are no further restrictions on the non-trivial metric functions
$H^{(0)}, W^{(0)}_i$).

{\em Acknowledgements: We would like to thank J.M.
Figueroa-O'Farrill for helpful comments and, in
particular, for pointing out the possibility of solutions with all 
RR fields. This work was supported
by NSERC (AC and NP), AARMS (SH) and the programme FP52 of the
Foundation for Research of Matter, FOM (AF). AF would like to
thank Dalhousie University for its hospitality while (part of)
this work was carried out, and L. Maoz for helpful discussions.

\end{document}